\documentstyle[12pt,moriond,psfig]{article}
\begin{document}
\heading{Bulge Formation}

\author{F. Combes }
{ Observatoire de Paris, DEMIRM, Paris, France}

\begin{moriondabstract}
  The currently discussed theories of bulge formation are reviewed,
including the primordial scenario, where bulges form rapidly and 
then accrete disks, the secular scenario, where bulges are formed by
dynamical evolution of disks through bars and galaxy interactions,
and some combinations of both, where formation of bulges and disks
are more continuous and interleaved. The various scenarios make specific
predictions about the relative masses, angular momenta, colours,
metallicities of bulges relative to disks, and the bulge-to-disk ratio
as a function of time.  Dynamical processes relevant to the formation
of bulges (bar instabilities, mergers) are described and tested against
observed statistics. Current data suggest a dynamical feedback from 
gravitational instabilities in bulge and disk formation.
It is very difficult to discriminate between the various scenarios from
surveys at $z=0$ only, and observations at high redshift 
are presently the best hope for large progress. 
\end{moriondabstract}

\section{Introduction}

The problem of bulge formation is in rapid evolution:
not only many scenarios, dynamical processus, formation theories
have been proposed and studied, but there has been great
progress in observation of bulges: on ages, metallicities,
structures, etc... One can cite in particular:

\begin{itemize}

\item Detailed age study of individual globular clusters, with colour-magnitude
diagrams of individual stars, with the high spatial resolution of HST
(in the Milky Way bulge, and in Local Group galaxies)

\item More extinction-free studies of bulge structure through near-infrared
imaging, made recently possible at large scale (wide extragalactic
surveys, COBE for the Milky Way)

\item Morphological studies of galaxies at high redshift: this
has been extensively developped in this meeting, and is the
privileged tool to tackle galaxy evolution in situ. 

\end{itemize}

There are very good recent reviews on the subject, namely in 
proceedings of the STSci workshop held in 1998 (``When and how bulges
form ?'', ed. Carollo, Fergusson \& Wyse),  Renzini (1999), Silk \& Bouwens
(1999), Carlberg (1999) or the Annual Review on ``Galactic Bulges'' by
Wyse, Gilmore \& Franx (1997).

The main scenarios proposed for the fornation of bulges are:

\begin{itemize}

\item  Monolithic formation (Eggen, Lynden-Bell \& Sandage 
 1962), or very early dissipative collapse at the beginning of galaxy
formation. This assumes that the gas experiences violent
3D star formation, so quickly that it had no time to
settle into a disk. This scenario was first proposed to explain
the almost spherical old metal-poor stellar halo 
and the subsequent formation of disks with different thickness,
that were thought to be an age sequence following the
progressive gas settling. This is now proposed for
the central bulge, the disk being acquired later.  

\item Secular dynamical evolution: the time-scale of these
processes are longer than the dynamical time (i.e. secular),
but however smaller than the Hubble time. They are due
to the dynamical interaction between the various components,
disk, bulge, halo. Gravitational instabilities, such as bars and
spirals, are able to transfer efficiently angular momentum, and
produce radial mass flows towards the center. Due to vertical
resonances, stars in the center  are elevated above the plane, 
and contribute to bulge formation (Combes et al. 1990).

\item galaxy interactions, through merger and mass accretion.
It is well known that major mergers between spirals can
result in the formation of an elliptical galaxy (Toomre \&
Toomre 1972). Similarly, the accretion by a spiral galaxy of
a dwarf companion could contribute to the formation of
a spheroid at the center, assumimg that this minor merger
has not destroyed the disk. 

\end{itemize}

In fact, all three of these main scenarii certainly occur, the
question is to estimate the relative role of each of them,
which is related to the time-scale of bulge formation.
All these processes may be included in the general frame
of hierachical galaxy formation. Disks are supposed to
form through gas cooling in a dark halo. Cooling runs
progressively from the center to the outer parts
(case of continuous gas infall), and produces an
inside-out formation of disks. Either this cooling is 
first violent at the center, due to the short dynamical
time-scales, and the absence of a stabilising heavy stellar 
object, and a spheroid can form first; or secular 
dynamical evolution could afterwards transfer angular
momentum, and bring mass to the center.
 In all scenarii, the speed of evolution and rate of
star formation depends strongly on environment.
Galaxy interactions are both responsible of the merger
scenario, and also trigger or boost bar formation
and secular evolution.
Bulges at the center of galaxies accumulate all stars
formed, and in any case they are expected to be older than
the outer disk, and mainly with much more scattered
properties (ages, abundances). 

\smallskip

We will first briefly review the constraints and clues
brought by observations related to bulge formation,
and then examine each scenario respectively.

\section{Clues from observations}

Observations of our own bulge is impeded by 
dust extinction, confusion (crowding), contamination
by foreground stars, that make data on the Milky Way very uncertain.
There exist in the literature  a certain number of 
prejudices, like the idea that ``bulges are old and metal-rich, 
and small versions of ellipticals'', that are not today
completely confirmed: the reality is not so simple, as clearly
reviewed by Wyse et al (1997).

\subsection{ Metallicity}

In the Milky Way, the mean metallicity of the bulge
is the same as that of the disk in the solar neighborhood,
but in contrast to the disk, the bulge has a very wide scatter.
This characteristic allows the metallicity distribution of
the bulge to be explained by a closed box chemical model,
contrary to the disk: the latter has a very narrow distribution
of metallicities, that gives rise to the well-known G-dwarf problem
(its solution requires other processes, like gas infall, etc..).
In the Andromeda bulge also, super metal-rich globular
clusters are the exception (Jablonka et al 1998).

In the Milky Way bulge, there is no correlation between age, color, abundance 
and kinematics, which could have given clues about the origins
(e.g. Rich 1997).
There does not seem to have been a starburst, since there
is no excess of $\alpha$ elements (Mc William \& Rich 1994).

\subsection {Age}

As for ages, however, we should not make the confusion:
{\it the bulge has not necessarily the age of its stars}.
In particular the bulge could have formed recently from
old stars. It is true namely if the bulge has formed from bars.
 Thanks to the clear overall structure of the MW given
by COBE-DIRBE, we know directly now that a bar is present,
 while it has long be assumed from gas kinematics only
(e.g. Peters 1975). The peanut/boxy bulge of the MW 
has asymmetries due to a bar seen in perspective
(e.g. Blitz \& Spergel 1991, Zhao et al 1996). It is even difficult to distinguish 
what is bar and what is bulge (Kuijken 1996). Since strong bars
are only a transient phase (see below) it is easy to extrapolate
how the stars presently in the bar will form the bulge.

\begin{figure}
\psfig{figure=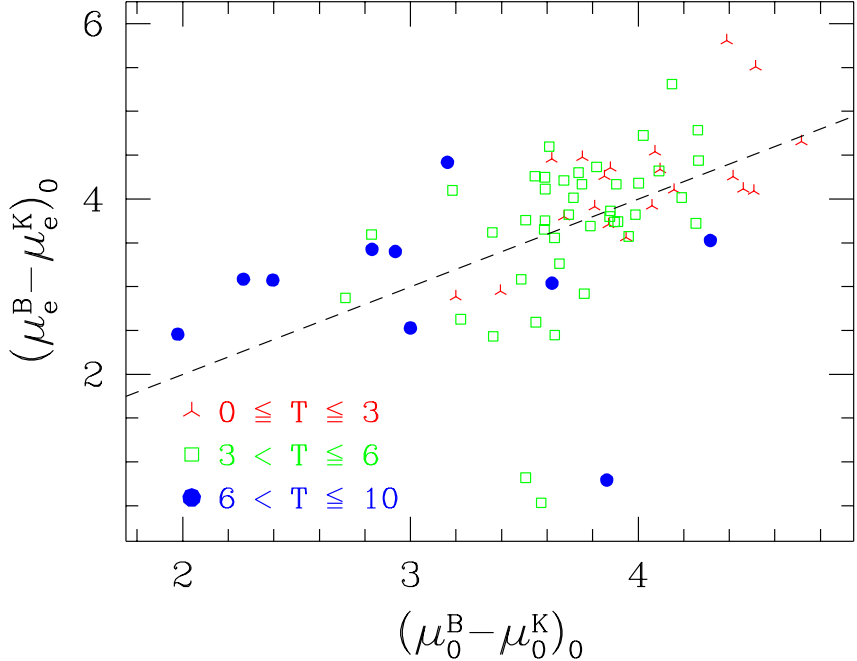,width=12cm,bbllx=1cm,bblly=5cm,bburx=19cm,bbury=20cm,angle=0}
\caption{The $B-K$ color in surface-brightness of the central disk, versus
that of the bulge, from de Jong (1996). The various symbols indicate morphological
types, as described at the bottom left. The dashed line means equality.   }
\label{fig1}
\end{figure}

If the bulges are thought to be older than disks, it might
be due to the implicit average over the whole disk.  
In fact, when bulges and inner disks are compared, they
have the same integrated colors (Peletier \& Balcells 1996, 
de Jong 1996, see fig \ref{fig1}). But there exist clear
   radial gradients of colors and metallicities.
Spiral galaxies become bluer with increasing radius.
Moreover, colors correlate well with surface brightness.
These colors and their gradients can best be explained 
with history of star formation, and dust reddening is
not dominant, as shown by de Jong (1996): 
outer parts are younger than central regions.
These observations support an inside-out galaxy formation.
Alternatively, dynamical secular evolution can
also account for these observations, through angular momentum transfer 
(cf Tsujimoto et al 1995).

\subsection{ Are bulges similar or not to giant ellipticals ?}

Bulges are spheroids that follow the same fundamental plane
as elliptical galaxies. They also follow the same 
luminosity-metallicity relation (Jablonka et al 1996).
But they are oblate rotaters (flattened by rotation),
while giant ellipticals are not rotating. However, 
 low-luminosity ellipticals also rotate (Davies et al 1983).
There might be a continuity between bulges, lenticulars
and ellipticals. Most of the latters have been observed 
with compact disks, e.g.  Rix \& White 1992, Rix et al. 1999).
 There is also continuity in the light profiles.
They can be fitted as an exponential of 
$r^{1/n}$, while $n$ is a function of luminosity 
($n=$4 for Ellipticals, $\infty$ for Bulges, Andredakis et al 1995).

\subsection{ High-redshift galaxies }

From the CFRS and LDSS, surveying galaxies at redshifts
below 1, there is very little evolution of the
luminosity function of red galaxies
(Lilly et al. 1995, 1998). Their
disk scale-length stays constant, up to $z=1$ (which
could mean either no-evolution, or stationary merging).
On the contrary, there is substantial evolution in the blue
objects. Does it mean that a 
large fraction of bulges/spheroids are already formed? 
(or mean age of stars $\propto$ age of Universe).
From HST images of the CFRS galaxies,
Shade et al (1995, 96) conclude that
red galaxies have large bulge-to-disks (B/D) ratios, and that
most blue galaxies are interacting (also forming bulges, nucleated
galaxies).

Steidel et al (1996) from their sample of z $\sim$ 3 galaxies
find in average objects smaller than today 
(see also Bouwens et al 1998).
If violent starbursts are forming at high redshift, like the
ultra-luminous IR galaxies observed nearby, they should
dominate the sub-mm surveys. Already a large 
fraction of the cosmic IR background has been resolved
into sources (Hughes et al. 1998). If these objects are
forming the spheroids (bulges and ellipticals), then their
epoch of formation is relatively recent $z < 2$ (Lilly et al. 1999).

From the high spatial resolution of HST, it becomes now
possible to study the morphology of high-redshift galaxies,
and address the evolution of the Hubble sequence.
At least, it is possible to determine the concentration
(C) and asymmetry (A) to classify galaxies (Abraham et al 1996, 99).
If there is a consensus among the various studies, it is
on the considerable increase of perturbed and interacting
galaxies ($\sim$ 40\% objects interacting).
But on the concentration, or bulge-to-disk ratios,
there is no unanimity. Abraham et al (1998) do not
find evolution in the B/D ratio, and 
tend to favour monolithic bulge formation.
Marleau \& Simard (1998) find on the contrary many
more disk-like faint galaxies in the Hubble Deep Fields, i.e.
that there are less objects with high bulge-to-disk ratios
at high redshift (with respect to $z=0$).

\section{Monolithic dissipative formation}

 This kind of violent and intense star formation 
is different to what is seen today in galactic disks: it
occurs in three-dimension (Elmegreen 1999).
In the deep potential well of a giant galaxy center,
almost entirely due to the dark halo, there cannot be
SF self-regulation by blow-out (supernovae, winds, pressure) 
as in disks or dwarfs today (Meurer et al 1997).
Most of the time, in the ultra-luminous galaxies
of the nearby universe, the starburst occurs in nuclear rings or disks 
(Downes \& Solomon 1998). This could be different in the 
early universe, since there are no heavy disks formed.
The threshold for SF in 3D can be estimated from the virial theorem
(Spitzer 1942) as a critical central volumic gas density
$\rho_c = \rho_{vir} /(1+\beta/2)$ for gaseous potential 
$\beta GM_{gas}/R$. The star formation rate SFR is then
$SFR \propto \rho_{vir}/t_{dyn} \propto  \rho_{vir}^{3/2}$ 
(an equivalent of Schmidt law). In localised dense regions, 
the SFR could be very large, and an
extremely clumpy distribution of stars 
is expected. Simulations illustrating this process
have been done by Noguchi (1998).

\section{Secular evolution}

\begin{figure}
\psfig{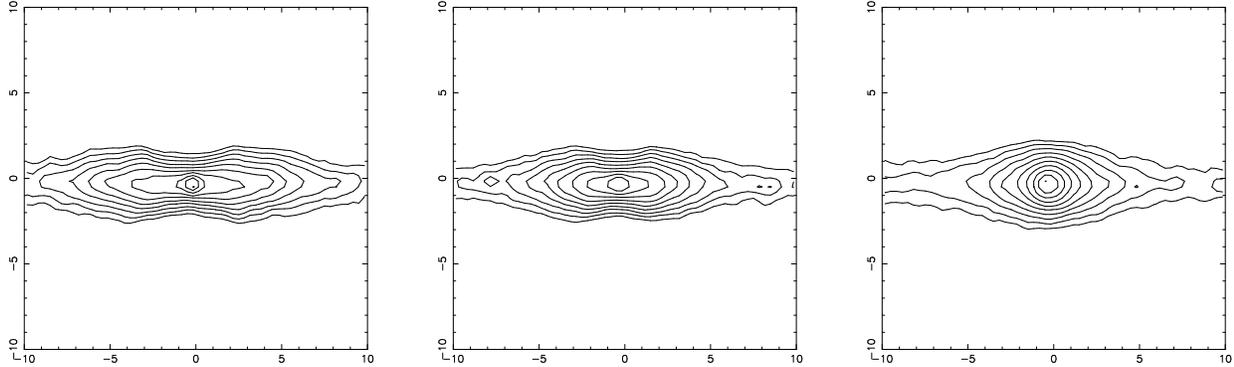}
\caption{Edge-on projections of a bar obtained in N-body simulations, for 
three orientations of the bar with respect to the line of sight (90, 45 and 0$^\circ$).}
\label{fig2}
\end{figure}

A wide series of N-body simulations, supported by
observations, have established the various phases
of this secular dynamical evolution
(e.g. Sellwood \& Wilkinson 1993, Pfenniger 1993,  
Martinet 1995, Buta \& Combes 1996). They can
be described as:

\begin{itemize}

\item Gravitational instabilities spontaneously
form bars, spirals, which transfer angular-momentum
through the disk. 
Both stars and gas lose momentum to the wave.
Radial gas flows in particular produce large
central mass concentration.

\item Due to vertical resonances, the bar thickens
in the center (box or peanut-shape, Combes et al 1990),
and contributes to bulge formation (see fig \ref{fig2}).
 This scenario is compatible with the observed
correlation between scale-lengths of bulges and disks  
(Courteau et al 1996).

\item Sufficient mass accumulation in the center 
(1-5\% of the total disk mass) perturbs the rotation curve,
changes the orbit precession rates, and 
destroy the bar, by creating perpendicular orbits 
$x2$. The galaxy has now become a hot stable system,
with a central spheroid.

\item Through gas infall,
the disk can become unstable again  and reform a bar 
(with a different pattern speed)

\end{itemize}

\section{Accretion and mergers}

Ellipticals can form by the mergers of two spirals 
(e.g. NGC 7252 prototype); but the most general case
is the formation by the merger of several smaller objects.
That ellipticals have accreted several smaller companions
during their life is supported by the frequency of shells 
(as much as 50\% according to Schweizer \& Seitzer 1988).
Bulges could similarly be the result of minor mergers;
even today, there exist numerous companions around giant spirals.

In hierachical cosmological models, it is easy to
compute analytically the history of dark halo formation,
and their merging rate (Kauffmann et al 1993, Baugh et al 1996).
But the fate of the dissipative visible matter is still
not well-known. According to some star formation and feedback
parameters that fit the colour-magnitude relation of ellipticals
in clusters,  Kauffmann \& Charlot (1998) have shown that
massive galaxies must have formed continuously, and 
must have assembled only recently,  
in order to fit observed redshift distributions of K-band luminosity at $z=1$.

If the bulge comes from accretion, observations of our own MW 
put constraints on the metallicity of the objects accreted 
(they should be relatively high).
This put strong limits on the fraction of the bulge that has been accreted  
recently (Unavane et al 1996).

\section{Combination of all scenarii}

In a hierachical scenario, there are many uncertainties
about the physical processes controlling  the 
baryonic matter. Feedback must prevent efficient conversion of 
gas into stars, to represent observations. There remains much freedom
to determine it more quantitatively, as well as the distribution of 
angular momentum, etc...

The simplest recipe is to assume that the angular momentum of
the matter is not redistributed, and that the gas cools at the 
radius where the cooling time becomes shorter than the dynamical time
(Mo et al 1998; van den Bosch 1999). This represents gas infall as
cooling flows in dark haloes, and the formation of disks as an 
inside-out process. There are however
many hypotheses possible: for instance,  
that the mass of the disk is a fixed fraction of the halo mass,
or that the angular momentum of the disk is a fixed fraction of the halo 
angular momentum; also that the 
disk is exponential in shape, and is stable (which gives a 
condition on specific momentum). The efficiency
of galaxy formation ($\epsilon_{gf}$) is also a free parameter, as
a function of redshift; it could vary proportionally to
the Hubble constant, to ensure the Tully-Fisher relation 
(van den Bosch 1999).
The specific angular momentum of the galaxies are supposed
to be acquired by tides (Fall \& Efstathiou 1980).

These scenarios, coupled with sufficient feedback (due to star-formation)
can explain high-surface brightness galaxies (HSB) and even LSB, 
if a range of angular momentum is assumed (Dalcanton et al 1997).
 The following constraints can be satisfied:
1) the Tully-Fisher scaling relation;  
2) the density-morphology relation (Dressler 1980);
3) the surface-density/size relation $\mu$ - R$_d$
(generalisation of Kormendy 1977).

A general feature found by van den Bosch (1999), whatever the cooling/feedback
hypotheses, is that the present observed disks lie very close to their
stability limit (with respect to strong gravitational instabilities), which suggests
a self-regulated formation, through dynamical processes:
the disk alone is unstable until a sufficient bulge stabilises it.
The gas infall could then alternatively favor bulge or disk formation
according to the bulge-to-disk mass ratio it encounters at a given
time. This could explain the observed coupling between
bulges and disks.  

\begin{figure}
\psfig{figure=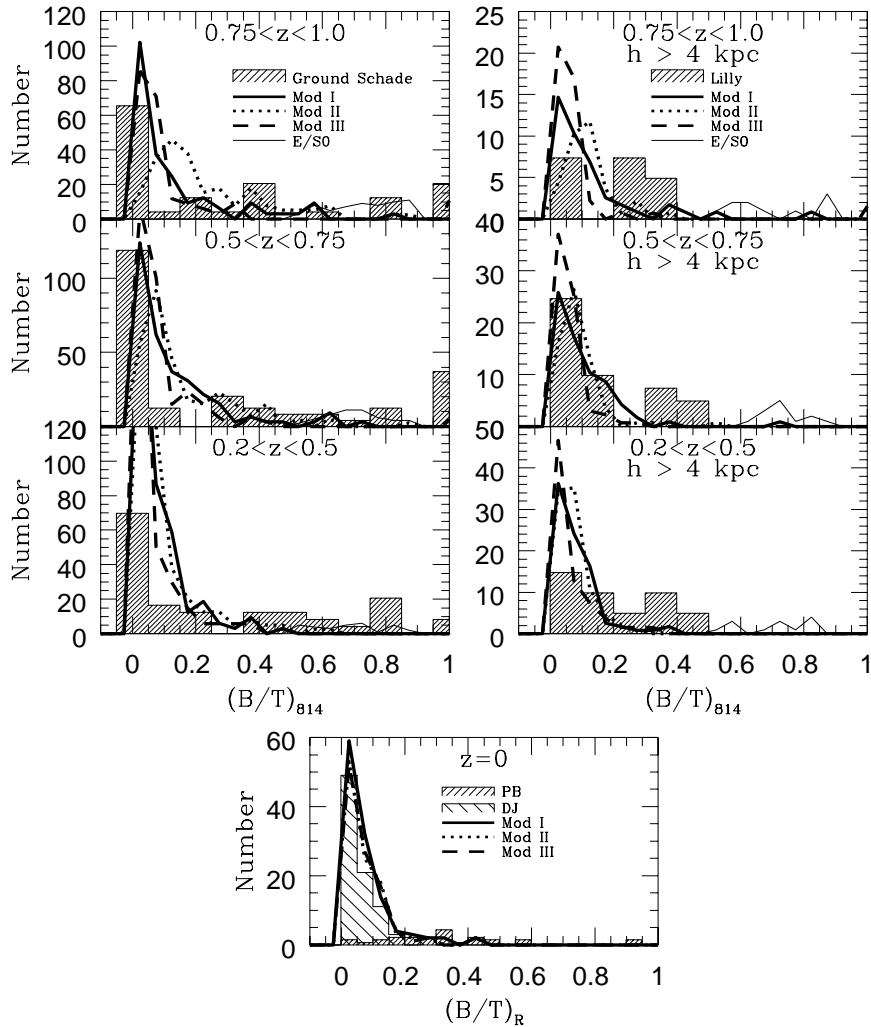,width=12cm,bbllx=1cm,bblly=45mm,bburx=17cm,bbury=24cm,angle=0}
\caption{ Comparison of the bulge-to-total ratios (B/T) between observations
(histograms) and model predictions (lines), from Bouwens et al (1999). At $z=0$,
the data are from Peletier \& Balcells (1996, PB) and de Jong (1996, DJ). 
At high $z$ from Schade et al (1996) and Lilly et al (1996). In all frames,
the models are: {\bf I} = secular evolution; {\bf II} = simultaneous formation; 
{\bf III} = early bulge formation. }
\label{fig3}
\end{figure}

Bouwens et al. (1999) have recently confronted several bulge formation scenarii
to the observations. At $z=0$, it appears almost impossible to distinguish
between early and late bulge formation (cf fig \ref{fig3}). At high redshift,
it becomes possible, essentially because a large fraction of the star formation
occurs at the observed $z$. For instance, in fig \ref{fig3} at top-left, the model
of  simultaneous formation (of disk and bulge) departs from the two others,
since star-formation makes the bulge momentarily brighter.

Could the star formation history (SFH) constrain the period
of formation of bulges and ellipticals? 
Since spheroids contain 30\% (Schechter \& Dressler 1987), 
or 66\% (Fukugita et al 1998) of the stars in the Universe,  
they cannot form too early, according to the Madau et al. (1996) 
SFH curve. However, this constraint is only a lower-limit
on the formation time,
since the spheroids could be recently formed from old stars.

\section{Conclusions}

Bulges are very heterogeneous structures, with 
large scatter in colors, metallicities and ages.
At large luminosities, bulges become similar to elliptical
galaxies, at low luminosities, they resemble more 
spiral disks (or bars). There exist objects with a whole
range of properties ensuring continuity between these
two extremes, suggesting evolution or the combination
of several processes.

Bulge formation is certainly a combination between
three main scenarii:
 
a fraction of bulges could have formed early (at first collapse);
then secular dynamical evolution enrich them; in parallel,
according to environment, accretion and minor mergers 
contribute to raise their mass.

Big spheroids (SOs, Es) can only form through major/minor mergers.
Present disks have been (re)-formed recently.
To be more quantitative, and precise the time-scale of bulge 
formation, a privileged solution is to observe and classify
the morphology of galaxies as a function of redshift, together
with detailed balance of the comoving volumic density of
galaxies of each type. This is necessary to avoid confusion between
a no-evolution model, at a given type, or a stationary evolution flow 
along the Hubble sequence.

\vfill
\end{document}